\documentclass[aps,prl,twocolumn,groupedaddress]{revtex4-1}

\usepackage{graphicx}
\usepackage[caption=false]{subfig}
\usepackage{hyperref}
\hypersetup{
colorlinks=true,
allcolors=blue
}

\begin{document}

%\preprint{}

\title{Mixing and de-mixing of model microswimmers in bi-motility mixtures}

\author{Adyant Agrawal}
\email[]{adyant2@gmail.com}
%\homepage[]{Your web page}\D\Delta \nuelta \nu
%\thanks{}
\affiliation{Department of Physics, Indian Institute of Technology, Delhi}
\author{Sujin B. Babu}
\email[]{sujin@physics.iitd.ac.in}
\affiliation{Department of Physics, Indian Institute of Technology, Delhi}

\date{\today}

\begin{abstract}
Cooperation between micro-organisms give rise to novel phenomena like clustering, swarming in suspension. We study the collective behavior of the artificial swimmer called Taylor line at low Reynolds number using multi-particle collision dynamics method. In this paper we have modeled bi-motility mixtures of multiple swimmers in $2$ dimension, which differ from each other by the velocity with which they swim. We observe that the swimmers can segregate into slower and faster ones depending on the relative difference in velocity of the $2$ type of swimmers. We also observe that  contribution of slower swimmers towards clustering, on an average, is much larger compared to faster ones, although we employ a homogeneous mixture. When the difference in velocity is large between the swimmers, the faster ones move away from the slower ones towards the boundary. On the other hand, when the relative difference in velocity is very small, the slower and faster swimmer mix together to form big clusters. At later time even for small difference in velocity the swimmers segregate into fast and slow swimmer clusters.
\end{abstract}


%\keywords{}
\maketitle

\setlength{\belowcaptionskip}{-2pt plus 2pt minus 0pt}
The collective motility of micro-organisms is quint-essential in a range of biological activities \citep{ Benisty2015, Sokolov2012,  Kearns2010, Hoeniger1964, Rauprich1996,  Wu2007, Lauga2009, Vicsek2012, Lauga2006, Heddergott2012, Babu2012a}. Large scale cooperative movement is seen in micro-organisms which propagate by virtue of deformations along the cell body. Such swimming strategies are commonly seen in spermatozoa, C. elegans and various flagellated microswimmers \citep{Lauga2006, Lauga2009, Vicsek2012, Oberholzer2010, Heddergott2012, Babu2012a, Vicsek1995, DiLuzio2005, Hu2015, Munch2016, Li2017,  Malgaretti2017, Hu2015a, Yang2010, Elgeti2014, Yang2008, Reigh2012, Damton2010, Peruani2012}. Micro-swimmers demonstrate various aggregating patterns such as swarming, clustering or band formation \citep{Hoeniger1964, Rauprich1996, Huepe2004, Peruani2006, Wu2007, Yang2008, Lauga2009, Kearns2010, Damton2010, Peruani2012, Sokolov2012, Vicsek2012, Zottl2014, Yang2014, Ariel2017, Stenhammar2017, Swiecicki2013, Copeland2009, Ramaswamy2010, Abkenar2013, Li2016, Elgeti2014, Blaschke2016, Zottl2016, Smrek2017, SRMcCandlish2012, Weber2016, Reigh2012}. Aggregation is a consequence of homology in the system and is hydro-dynamically favorable as it reduces energy consumption in transport \citep{Yang2008}. The real systems comprise of swimmers having a range of different motilities. In addition, a portion of swimmers may be unhealthy or may employ atypical swimming strategies and hence immensely differ in propagating strengths. To study how the cluster formation is favored in case when the species are not perfectly homologous, several simulations have been performed \citep{Elgeti2014, Yang2008, Yang2010, SRMcCandlish2012, Awazu2014, Stenhammar2015, Weber2016, You2009}. Consequently, a positive feedback between clustering and segregation has been reported \citep{SRMcCandlish2012, Vicsek2012}. Previous works comprising mixture of active and passive self-propelled particles or rods indicate spontaneous segregation in the system \citep{SRMcCandlish2012, Awazu2014, Stenhammar2015, Weber2016, Smrek2017}.
These processes occur even in absence of cell to cell signaling or chemotaxis \citep{Damton2010, Yang2008, Lauga2009, Copeland2009, You2009, Kearns2010, Ramaswamy2010, Yang2010, Vicsek2012, SRMcCandlish2012, Abkenar2013, Weber2016, Li2016, Li2017}.
%A description on group dynamics leading to aggregation in such species and the arrangement inside a cluster have been aspired for.

In this Letter, we employ numerical simulations to investigate the collective dynamics of microswimmers which show propulsion via planar beating mechanisms in Newtonian fluid. We analyze the cooperation between the swimmers in a bi-motility mixture which results into the aggregation and segregation among their own types. Understanding this cooperation is prerequisite to deeper understanding of collective motion of microswimmers. In the present study, we consider only hydrodynamic and steric interactions between the swimmers. To model artificial swimmers, we use a two dimensional discretized model of Taylor’s sheets termed as Taylor Lines \citep{Munch2016}. The Taylor Line hydrodynamically interacts with the fluid using a sinusoidal bending wave which moves along the body. To simulate fluid environment, Multi-Particle Collision Dynamics (MPC) is used which employs “coarse-grained” particles of mass $m=1$ \citep{Malevanets1999, Lamura2001, Gompper2009}.
% The typical execution of MPC may result in unbalanced moment in fluid as well as may induce heating of the system as the Taylor line is continuously pumping energy into the system.

As the Taylor line is continuously pumping energy into the system, we have used MPC with Anderson Thermostat and angular momentum conservation (MPC-AT+a) \citep{Gompper2009}. The method consists of consecutive collision and streaming steps. In ballistic streaming step, the coordinates of the fluid particles $\{\overrightarrow{r_i}\}$ having velocity $\{\overrightarrow{v_i}\}$ are updated with integration time Dt.
%\begin{equation}
%\overrightarrow{r_i}(t+Dt)= \overrightarrow{r_i}(t) + Dt\cdot\overrightarrow{v_i}(t)\label{eq0}
%\end{equation}
For the collision step, the particles are segregated into collision cells of length $l=1$ and are imparted random velocities chosen from the Gaussian distribution of variance $k_B T/m=1$ such that the momentum of the cell is conserved, where $k_B$ is Boltzmann constant and $T$ is the temperature. The position of grid with respect to box is randomly changed in each step so as to incorporate Galilean invariance.
%The system parameters in our simulations are measured such that unit of energy $[E]=k_B T/m$, unit of length $[L]=l$ and mass has unit of $m$. Thus the unit of time $[T]=l\sqrt{(m/k_B T)}$.
We have chosen the density of fluid as $10 m/l^3$. The unit of time is $[t]=l\sqrt{(m/k_B T)}$.

Each Taylor line consists of a sequence of N beads each of mass 10m at an equilibrium distance of $\frac{1}{2} l$. These beads interact with the nearest neighbors by Hooke's spring potential and bending potential $V_B=\frac{\kappa}{2} \sum_{i=1}^{N-1}[\overrightarrow{t}_{i+1}-R(\alpha_i)\overrightarrow{t_i}]$ \citep{Munch2016} that keeps the consecutive beads aligned at an angle $\alpha_i$ with the bending rigidity $\kappa=pk_B T$ where $p$ is the persistent length. A sinusoidal wave of beating frequency $\nu$ is generated along the contour of the Taylor line by varying curvature $c(i,t)=\alpha_i/l$, spontaneously, with time, $t$ and bead position, $i$.
\begin{equation}
c(i,t)=b\cdot \sin \left[ 2\pi \left( \nu t + \frac{2i}{N} \right) + \phi\right] \label{eq1}
\end{equation}
To get a directed motion we choose $p=5\times10^3 \cdot l (N-1)$. The factor of $2$ with $i$ assures that there are two waves trains packed inside the Taylor line and $\phi$ is the the initial phase shift. In order to model the interaction among various swimmers, we use a truncated Leonard-Jones potential,
\begin{equation}
V_I = 4\epsilon \left[\left(\frac{r_o}{r}\right)^{12} - \left(\frac{r_o}{r}\right)^6 \right] ,\,\,\,\,\,\,\,\, r^2<2^3 r_o\
\end{equation}
where $r$ is the separation between the beads of different Taylor lines, $r_o$ is  taken to be equal to $l$ and $\epsilon$=13.75 \citep{Yang2008,Yang2010} is the strength of the potential. Using this potential along with the intra-swimmer potentials, we calculate the acceleration on every bead and update their position with integration time step $dt = 0.01Dt$. To simulate the hydrodynamic interaction between the swimmer and the fluid, we make the beads to participate in every collision step.
% that occurs once in every $100 dt$ time steps. 
This ultimately incorporates the Taylor line into the fluid environment.

We make use of both rigid and periodic boundary walls in this study. In case of rigid boundary, we use $L\times L$ square and a circle of radius $R$ as the kinds of wall.
% and in case of periodic boundary we use a $L\times L$ square. 
To mimic fluid flow at the rigid walls of a confinement, we bounce back the fluid particles crossing the walls and employ “Ghost” particle method in order to satisfy no-slip (rigid) boundary condition \citep{Lamura2001,Gompper2009}. We allow the Taylor lines to easily slide on the walls by implementing bounce forward rule on the beads. For the purpose of our simulations we choose $Dt = 0.01$, $N$ to be 100, and $b$ to be 0.15 which gives the equilibrium wavelength of 22.59 and amplitude 2.27.
% Frequency, $\nu$ in eq.(\ref{eq1}) is taken in a range of 0.001 - 0.009. 
A single swimmer in periodic boundary condition, yields a velocity of 0.0025 - 0.0224 for the respective frequency range of 0.001 - 0.009 similar to the work of \citeauthor{Munch2016} and the Reynolds number in the range of 0.003 - 0.028.

%Previous works \citep{Yang2008,Yang2010} have illustrated that in the process of propulsion, the energy consumption per unit time by a single swimmer is proportional to the swimmers beating frequency.
\citeauthor{Yang2010} showed that if the velocity of each swimmer was chosen from a Gaussian distribution, the cooperation of swimmers was enhanced when the variance, $\sigma<3\%$. In the present study we will analyze a simple system consisting of two types of swimmers which differ only in the swimming velocity. Since for Taylor line the velocity is directly proportional to the beating frequency $\nu$ \cite{Munch2016}, we vary the frequency of actuation in our simulations. The beating frequency of the faster swimmer is given by $\nu_a$ and that of slower swimmer is given by $\nu_b$ where we define $\delta \nu = |\nu_a-\nu_b|/\langle \nu \rangle$. The range of $\delta \nu$ in the present study is  $10^{-2} - 1$. We have performed simulations with a number density of the swimmers ranging from $\rho= 1.5 \times 10^{-3}$. Initially, the swimmers are scattered inside MPC fluid with random center of mass coordinate, orientation, initial phase and direction of motion.

\begin{figure}[t]
\centering
%\captionsetup[subfigure]{labelformat=empty}
\subfloat{(a)
\label{fig1.1}
\includegraphics[width=0.54\linewidth]{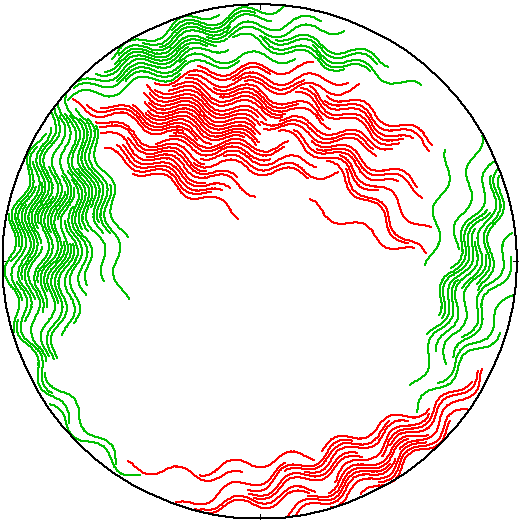}%
}%\hfill%
\begin{minipage}[b]{0.38\linewidth}
\subfloat{(b)
\label{fig1.2}%
\includegraphics[width=0.68\linewidth]{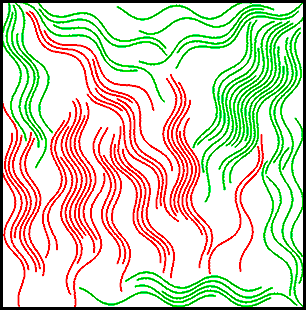}%
}%\hfill%\[\baselineskip]
\hfill\\[-1pt]
\subfloat{(c)
\label{fig1.3}%
\includegraphics[width=0.68\linewidth]{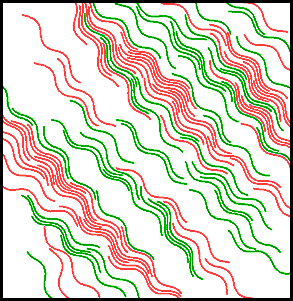}%
}%\hfill%
\end{minipage}
\caption{Snapshots of stable state of system with 3 kinds of boundaries evolved from a random initial state. (a) Circular confinement of $R=100$ with 150 swimmers, (b) $100\times100$ square confinement with 75 swimmers, (c) $150\times150$ periodic boundary with 110 swimmers. Each system contains uniform mixture of swimmers with 2 different beating frequencies ($\delta \nu$ = 0.1), $\nu_a$ = 0.00525 (green) and $\nu_b$ = 0.00475 (red).}
\label{fig1}%
\end{figure}

In Fig.\ref{fig1} we have shown the snapshots as obtained from the simulation for three different boundary conditions, where red color signifies slow swimmer and green signifies fast swimmer. In Fig.\ref{fig1.1} we show the aggregates formed due to circular rigid boundary condition, where we observe that the slow swimmers are usually along the center. A visual inspection reveals that the swimmers have segregated into slow and fast swimmer clusters. In Fig.\ref{fig1.2} we have shown the aggregate formed due to rigid square boundary condition, where we observe that the faster swimmers are forming cluster at the corners of the square. Here too a segregation between fast and slow swimmers can be seen. Supplemental Material\cite{suppl}, Fig.S1 shows the snapshots of the system at different time instants. While in Fig.\ref{fig1.3} we have used periodic boundary condition and observe that the segregation is accompanied by the formation of bands because boundaries do not exit in this case as reported by .... et.al. All the results we discuss from hereon are for rigid circular boundary condition only.
%As the system evolves, the swimmers form aggregates, clusters (in case of confinement) or bands (in case of periodic boundary) via steric interactions through potential and/or hydrodynamic interactions with other swimmers [Fig. \ref{fig1}]

To quantify the collective behavior, we calculate the cluster size of the swimmer as follows. We consider two swimmers to be part of a cluster if they simultaneously satisfy two conditions for one complete beating period of the faster swimmer. First, if the minimum distance between at least $10\%$ of the beads of the two swimmers is less than the 2.27a, which is the amplitude of swimmer in our case. Second, if the angle between the end-to-end vectors \citep{Munch2016}, is less than $\pi/6$. See Supplemental Material\cite{suppl}, Fig.S2 for an illustration of clusters.
% by our definition.

\begin{figure}[ht]
\centering
\subfloat{(a)\label{fig2.1}%
\includegraphics[width=0.95\linewidth]{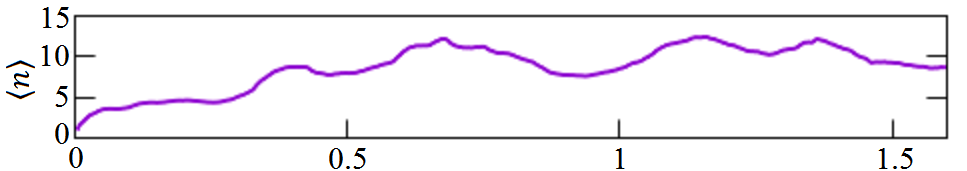}%
}
\hfill\\[-1pt]
\subfloat{(b)\label{fig2.2}%
\includegraphics[width=0.95\linewidth]{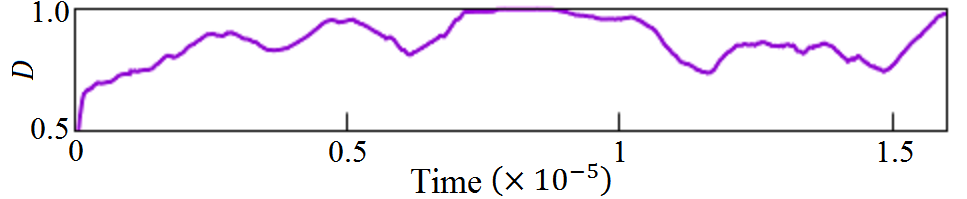}%
}
\hfill\\[-1pt]
\subfloat[t=$0.75\times10^{5}$ Mixed]{\label{fig2.3}
\includegraphics[width=0.25\linewidth]{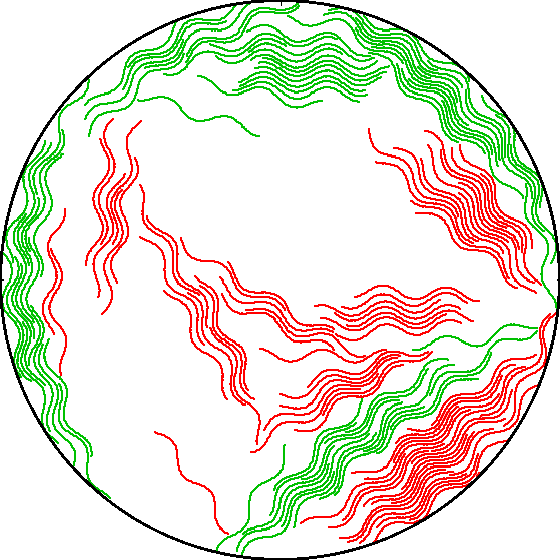}%
}\hfill
\subfloat[t=$0.85\times10^{5}$ De-mixed]{\label{fig2.4}%
\includegraphics[width=0.25\linewidth]{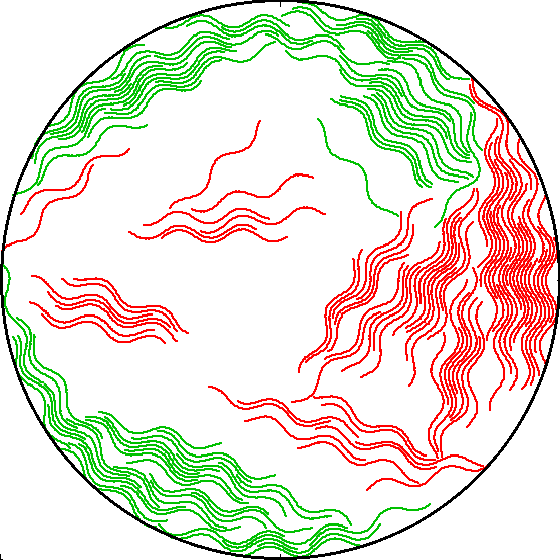}%
}\hfill
\subfloat[t=$1.05\times10^{5}$ Mixed]{\label{fig2.5}%
\includegraphics[width=0.25\linewidth]{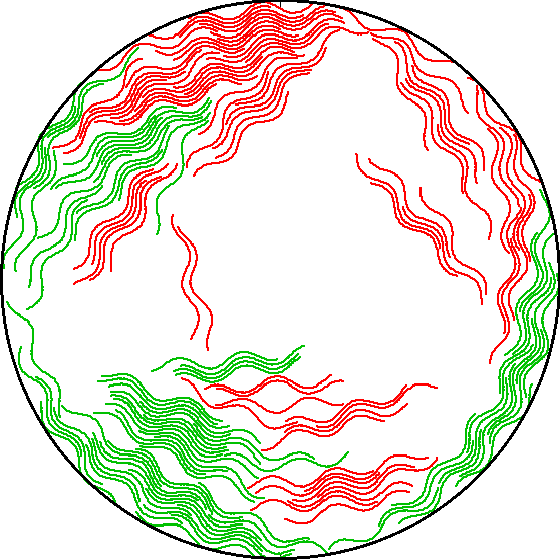}%
}
\caption{(a) Average cluster size and (b) segregation index ($D$) vs time for a system with circular confinement with radius R=100. Relative difference in frequency, $\delta\nu=0.4$ and average frequency, $\langle \nu \rangle = 0.005$. (Bottom) Snapshots of the system showing consecutive mixed, de-mixed and mixed states at t=$0.75\times10^{5}$,t=$0.85\times10^{5}$ and t=$1.05\times10^{5}$ respectively.}%
\label{fig2}%
\end{figure}

In Fig\ref{fig2.1} we show the evolution of the average cluster size, ${\langle n \rangle= \{\sum n\cdot\Pi(n)\}/{\sum \Pi(n)}}$, where $\Pi(n)$ is the cluster size distribution. We observe that initially $\langle n \rangle$ increases, signifying that clusters are being formed. These clusters keeps on growing un-till $\langle n \rangle$ reaches a steady state and then it oscillates around $10$ indicating there is constant aggregation and fragmentation of the clusters. To understand whether we have a segregated state or a mixed state we employ a dimensional number called the segregation index ($D$) \citep{Duncan1955}.
\begin{equation}
D = \frac{1}{2} \sum_{clusters} \left| \frac{n_a}{N_a}-\frac{n_b}{N_b}\right|
\end{equation}
where, $n_{a,b}$ is the population of $a$ or $b$ type swimmers in a particular cluster and $N_{a,b}$ is the total number of $a$ or $b$ type swimmers in the system. The summation runs over all the clusters in the system, which means $D=1$ implies completely segregated and $D=0$ implies fully mixed system. Fig.\ref{fig2.3} indicates $D$ initially increases and reaches $1$ signifying a completely segregated state and at later time it oscillates between $0.8$ and $1$. Also, we observe that when $\langle n \rangle$ is at a maximum $D$ is at a minimum and vice versa. In Fig.\ref{fig2.3} we have shown the snapshot of the system when we have a maximum in the system for $\langle n \rangle$ and a minimum in $D$ and we observe a mixed state where a cluster contains both fast and slow swimmer. Taylor lines always go towards the wall and form clusters close to the wall, these clusters then start moving along the walls. Even when the system is segregated both the types of swimmers would eventually encounter each other as one is slower than the other and will form mixed state. At $t=0.85 \times 10^{5}$ the mixed swimmers start to de-mix and segregate into clusters of fast and slow swimmers. At this time we observe that $D=1$ and $\langle n \rangle$ is a minimum, which means the swimmer have completely segregated as can be observed from the snapshot Fig.\ref{fig2.4}. Again at $t=1.05 \times 10^{5}$ we observe that the swimmers again mix together and form larger cluster as the $\langle n \rangle$ is a maximum and $D$ is a minimum, i.e when segregation is a minimum we always have large cluster on average in our system.

\begin{figure}[ht]
\subfloat[]{\label{fig3.1}%
\includegraphics[width=0.55\linewidth]{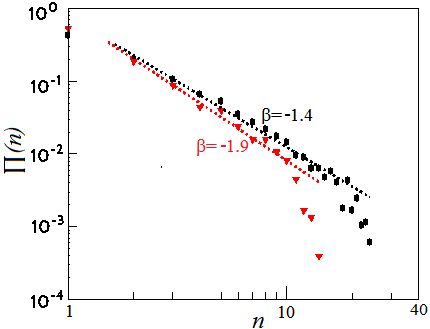}%
}\hfill%
\subfloat[]{\label{fig3.2}%
\includegraphics[width=0.45\linewidth]{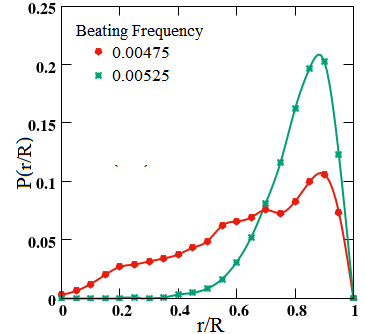}%
}
\caption{(a) Time averaged cluster size distribution for two different systems S1(red) and S2(black). S1 contains 100 swimmers in $200\times200$ square confinement while S2 contains 150 swimmers in circular confinement of $R = 100$. Density, $\rho_{S1}$=0.0025 \& $\rho_{S1}$=0.0048. The curves follow power law decay which breaks down at large values. The data is taken over 12 simulations for various frequencies of swimmers with $\delta\nu$ ranging from 0.005 to 1. (b) Probability of finding a swimmer at a fractional distance $r/R$ form center, P($r/R$) vs $r/R$ for a circularly confined system of $R=100$ with $\delta\nu = 0.1$}%
\label{fig3}%
\end{figure}

The cluster size distribution $\Pi(n)$ in case of circular confinement is plotted in Fig. \ref{fig3.1}. A power law decay, $\Pi(n) \propto n^{\beta}$, is observed for smaller cluster followed by an exponential decay for the higher values of $n$. The power law exponent $\beta$ is independent of $\delta\nu$ and depends only on the density of swimmers. Fig. \ref{fig3.1} shows $\beta$ is approximately $-1.4$ for high density ($\rho=0.0048$) and approximately $-1.9$ for low density ($\rho=0.0025$) similar to what has been reported before for self propelled rods, both in case of simulations \citep{Huepe2004, Yang2008, Yang2010} as well as experiments \cite{Peruani2006}. $\beta$ also does not depend on the time up to which the average is done, confirming that it is an inherent property of system. The decrease in exponent with density shows the importance of interactions for cluster formation. In Fig \ref{fig3.2} we have plotted the probability of finding a slow or fast swimmer at a distance $r/R$ from the center of the circle, when the system has attained a steady state. Here we notice that the fast swimmers reach the wall earlier and stay near the wall as the distribution is nearly zero towards the center while has a sharp peak close to the walls. In the case of the slow swimmers the distribution is very broad, with a very small peak close to the walls. We know that the Taylor lines prefer to be closer to the walls, but in the case of slower swimmers when they reaches the wall the faster swimmer swim through the cluster and fragments the slow moving cluster. The fragmented segment again move towards the center. See Supplemental Movie.1 \cite{suppl}.

\begin{figure}[t]
\centering
\subfloat{(a)\label{fig4.1}%
\includegraphics[width=0.9\linewidth]{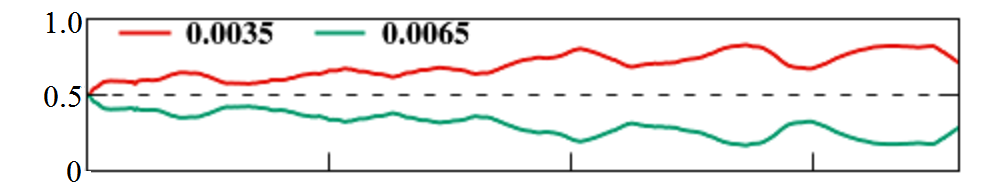}%
}
\hfill\\[-1pt]
\subfloat{(b)\label{fig4.2}%
\includegraphics[width=0.9\linewidth]{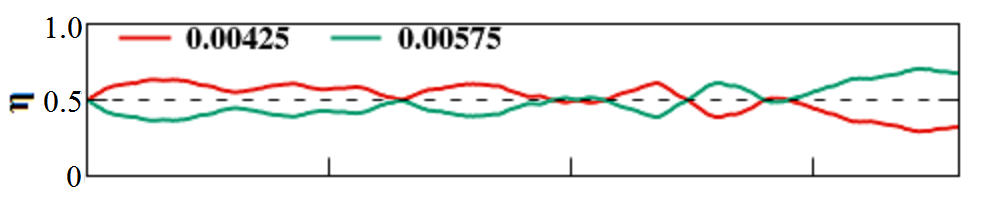}%
}
\hfill\\[-1pt]
\subfloat{(c)\label{fig4.3}%
\includegraphics[width=0.9\linewidth]{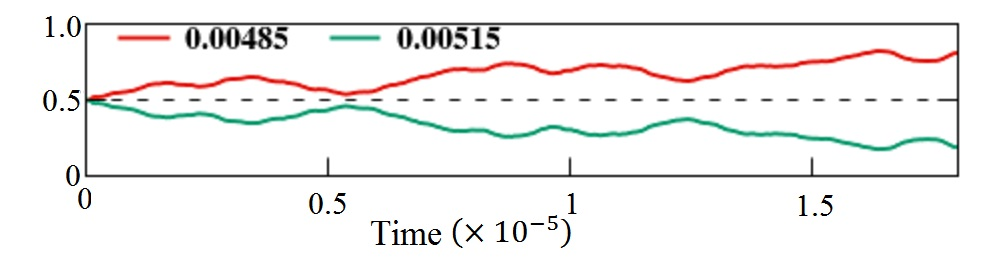}%
}
\caption{$\eta_a$ and $\eta_b$ vs time for three systems having circular confinement with R=100. For all systems, $N_a=N_b=75$. Swimmers are uniformly distributed at t=0. The systems differ in the beating frequency of swimmers, (a) $\delta\nu_1=0.6$, (b) $\delta\nu_{2}=0.3$ and (c) $\delta\nu_{3}=0.06$. The $\langle \nu \rangle$ in all 3 cases is 0.005. The segregation index for these systems is plotted in Supplemental Material \cite{suppl} Fig. S3.}%
\label{fig4}%
\end{figure}

The power laws suggest intermittent behavior in cluster dynamics \citep{Huepe2004} resulting from aggregation and fragmentation at steady state. To quantify the contribution of the fast and slow swimmer towards cluster formation we introduce a parameter $\eta$ defined as
\begin{equation}
\eta_{a,b}= \frac{\sum \left(\frac{n_{a,b}}{n}\right)\cdot\Pi(n)}{\sum \Pi(n)}
%\eta_{a,b}=\sum_{n_{a,b}} \frac{\Pi(n_{a,b})\cdot n_{a,b}}{\langle n \rangle}
\end{equation}
which give the contribution of either $a$ or $b$ swimmer in a cluster of size $n$. The time dependence of $\eta$ is plotted in Fig. \ref{fig4} for time period before the system reaches a steady state. We can observe that the contribution of slower swimmers to clusters is higher than that of the faster swimmers.In Fig. \ref{fig4} we have plotted the evolution of $\eta(a)$ and $\eta(b)$ for 3 different values of $\delta \nu$. It can be observed that for both large and small values of $\delta \nu$, the values of $\eta_b$ is always above the 0.5 and $\eta_a$ is below $0.5$. For the large difference in $\delta \nu$ the slower swimmers contribute more for the cluster formation, while faster swimmer prefer to form smaller cluster. As $\delta \nu$ is reduced, i.e., $\delta \nu=0.2$, $\eta_a$ and $\eta_b$ fluctuate around 0.5, where by the contribution towards the cluster by both the swimmers are almost the same. When the $\delta \nu$ is small, we again observe that the contribution of the slower swimmer are much more than the faster swimmers. If we further decrease the $\delta \nu$ we will observe that the contribution of both the kinds is the same.

\begin{figure}[t]
\centering
\includegraphics[width=0.85\linewidth]{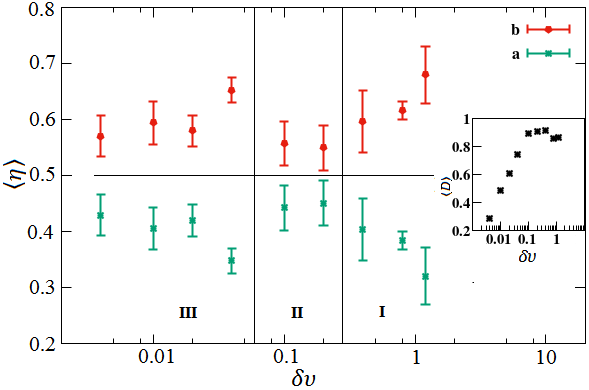}\hfill%
\caption{The time average of $\eta$ plotted vs. $\delta \nu$. The data is averaged over systems with circular confinement varying in $\rho$, R and $\langle \nu \rangle$. The inset shows the time average of segregation index vs.  $\delta \nu$. The average is taken over the simulation period.}
\label{fig5}%
\end{figure}

In the present work thus we are able to observe $3$ different regions based on the clustering of swimmers. Fig. \ref{fig5} shows a time average of $\langle {\eta} \rangle$ for the whole simulation period vs. $\delta \nu$. Each of the point is averaged over 6 different configurations with $\rho\approx0.005$. In region I, the faster ones push through the slower ones to reach the walls in small clusters while the slower ones are at that time dispersed around the center of the system. As a result, segregation index, as apparent from inset of Fig. \ref{fig5} and Supplemental Material \cite{suppl} Fig. S3, is greater than 0.8 and, $\langle \eta_b \rangle>0.5$ whereby the suspended slower ones will easily form clusters at center of the circular confinement. With the decrease in difference in beating frequency of swimmers in region II, the fast swimmers are unable to push through and there is virtually a competition between both the kinds of swimmers to form clusters in the confinement. As a result, there is almost, on an average, equal contribution from both the kinds to cluster and thus $\langle \eta_b \rangle$ tends to 0.5. If the $\delta \nu$ is decreased further in region III, both kinds of swimmers easily form clusters with each other as $\langle D \rangle <0.8$. $\langle \eta_b \rangle>0.5$ suggests that the concentration of slower swimmers is higher in a cluster and also the slow ones exploit the thrust of fast swimmers to form clusters. Thus the faster swimmers are always leading inside a cluster while there is a high density of slower swimmers at the back of the cluster. See Supplemental Movie.2 \cite{suppl}.
Gradually, the faster swimmers start swimming out of the cluster, thereby increasing the segregation. Thus, the parameter $\delta \nu$ plays an important role in controlling  the cluster dynamics of system. We have also simulated systems with $\delta \nu = 0.004$ to observe that as $\delta \nu$ tends to zero, $\eta$ and $\langle D \rangle$ tend to 0.5 and zero respectively as expected.

In conclusion, through our numerical simulation of hydrodynamically interacting Taylor's Lines in a confinement, we have shown that the cooperation between swimmers in a bi-motility mixture comprises of recurring mixing and de-mixing, which result in aggregation and finally segregation of the swimmer into fast and slow swimmer. This tendency of segregation has been reported in experiments \citep{Kearns2010,Ramaswamy2010,Damton2010,Benisty2015} in which such binary mixtures are developed artificially or in natural response to external stimuli and also in recent simulations of active and passive particles \citep{Awazu2014,Stenhammar2015,Weber2016,Smrek2017}. However, we have shown that the system shows different behaviors depending on the relative difference in velocity. The results can be exploited to understand the collective dynamics among microswimmers in real systems which are composed of a continuous distribution of motility. We can infer that a stable cluster of swimmers comprises of those with small difference in $\nu$, in which the slower ones are at the back guided by small numbers of faster ones, which is also observed experimentally \cite{Aureli2017, Vicsek2012}. When the difference in $\nu$ between clusters is large, the faster ones move away from the center assisting efficient swarming which has also been reported in the study of mixtures of healthy and dying microorganisms \citep{Benisty2015, Damton2010, Kearns2010}. Our simulations reveal the novel kinds of cooperation between different microswimmers which stimulate the collective motion in a suspension.

\begin{acknowledgments}
We would like to acknowledge HPC cluster at IIT Delhi, as well as baadal, the IITD’s private cloud, for allowing us to use the facility for the purpose of running the simulations.
\end{acknowledgments}

\bibliographystyle{apsrev4-1}
\bibliography{refer}

\end{document}